\documentclass[a4paper,reprint,aip]{revtex4-1}

\pdfoutput=0
\usepackage[T1]{fontenc} 
\usepackage{graphicx} 
\usepackage[USenglish]{babel} 
\usepackage{amssymb}
\usepackage{amsmath}
\usepackage{xspace} 
\usepackage{siunitx} 

\newcommand{\subs}[1]{${}_{\text{#1}}$}
\newcommand{\NCO}{Nd\subs{2}CuO\subs{4}\xspace}
\newcommand{\YBCO}{YBa\subs{2}Cu\subs{3}O\subs{7-x}\xspace}
\newcommand{\STO}{SrTiO\subs{3}\xspace}

\newcommand{\LSAT}{[LaAlO\subs{3}]\subs{0.3}[Sr\subs{2}AlTaO\subs{6}]\subs{0.7}\xspace}
\newcommand{\CNO}{(Nd,Ce)\subs{2}O\subs{3}\xspace}
\newcommand{\NCCOopt}{Nd\subs{1.85}Ce\subs{0.15}CuO\subs{4}\xspace}
\newcommand{\etal}{\textit{et~al.}\xspace}
\newcommand{\caxis}{{\it c}-axis\xspace}

\newcommand{\LSCOopt}{La\subs{1.85}Sr\subs{0.15}CuO\subs{4}\xspace}
\newcommand{\Tc}{{\em T\subs{c}}\xspace}

\newcommand{\miller}[3]{(#1\,#2\,#3)\xspace}
\newcommand{\millerx}[3]{[#1\,#2\,#3]\xspace}
\newcommand{\thth}{$\theta-2\theta$\xspace}

\newcommand{\mum}{\micro\meter}
\newcommand{\nm}{\nano\meter}
\newcommand{\mm}{\milli\meter}

\newcommand{\eV}{\electronvolt}

\begin{document}

\title{Strain accommodation through facet matching in \LSCOopt/\NCCOopt ramp-edge junctions}
\author{M. Hoek}
\altaffiliation{Currently at Boston College, 140 Commonwealth Ave, Chestnut Hill, MA 02467, USA.}
\affiliation{MESA+ Institute for Nanotechnology, University of Twente, P.O. Box 217, 7500 AE Enschede, The Netherlands}

\author{F. Coneri}
\affiliation{MESA+ Institute for Nanotechnology, University of Twente, P.O. Box 217, 7500 AE Enschede, The Netherlands}
\author{N. Poccia}
\affiliation{MESA+ Institute for Nanotechnology, University of Twente, P.O. Box 217, 7500 AE Enschede, The Netherlands}
\author{X. Renshaw Wang}
\altaffiliation{Currently at MIT, 77 Massachusetts Ave, Cambridge, MA 02139, USA.}
\affiliation{MESA+ Institute for Nanotechnology, University of Twente, P.O. Box 217, 7500 AE Enschede, The Netherlands}
\author{X. Ke}
\affiliation{Electron Microscopy for Materials Science (EMAT), Department of Physics, University of Antwerp, Groenenborgerlaan 171, B-2020 Antwerp, Belgium}
\author{G. Van Tendeloo}
\affiliation{Electron Microscopy for Materials Science (EMAT), Department of Physics, University of Antwerp, Groenenborgerlaan 171, B-2020 Antwerp, Belgium}
\author{H. Hilgenkamp}
\email[Author to whom correspondence should be addressed. Electronic mail: ]{h.hilgenkamp@utwente.nl.}
\affiliation{MESA+ Institute for Nanotechnology, University of Twente, P.O. Box 217, 7500 AE Enschede, The Netherlands}
\begin{abstract}
Scanning nano-focused X-ray diffraction (nXRD) and high-angle annular dark-field scanning transmission electron microscopy (HAADF-STEM) are used to investigate the crystal structure of ramp-edge junctions between superconducting electron-doped \NCCOopt and superconducting hole-doped \LSCOopt thin films, the latter being the top layer. 
On the ramp, a new growth mode of \LSCOopt with a \SI{3.3}{\degree} tilt of the \textit{c}-axis is found. 
We explain the tilt by developing a strain accommodation model that relies on facet matching, dictated by the ramp angle, indicating that a coherent domain boundary is formed at the interface. 
The possible implications of this growth mode for the creation of artificial domains in morphotropic materials are discussed.
\end{abstract}

\maketitle

Ramp-edge junction technology is an integral part of the research into the properties of (high-\Tc) superconductors \cite{Gao1990}. 
The ramp-edge configuration provides a good platform to create \textit{ab}-plane junctions between high-\Tc superconductors and between high-\Tc and conventional superconductors \cite{Hilgenkamp2002, Tafuri2005}. 
Additionally, ramp-edges are also used to create junctions through graphoepitaxy, where a large lattice mismatch between substrate and film causes the film to grow following the surface normal instead of the crystal direction of the substrate\cite{Faley2013a}. 
In this Letter, we show that there can be a third kind of junction that is not quite a full \textit{ab}-plane contact and where strain does play a role in determining the details of the growth, but not to the extend of promoting graphoepitaxy. 
We find that for appropriate lattice mismatches, a tilted phase can form on the ramp with respect to the \caxis aligned phase that grows away from the ramp; the tilting is dictated by facet matching to the lattice planes exposed on the ramp-edge, and is proportional to the ramp angle.

\begin{figure}[t]
\includegraphics[width =0.85 \columnwidth]{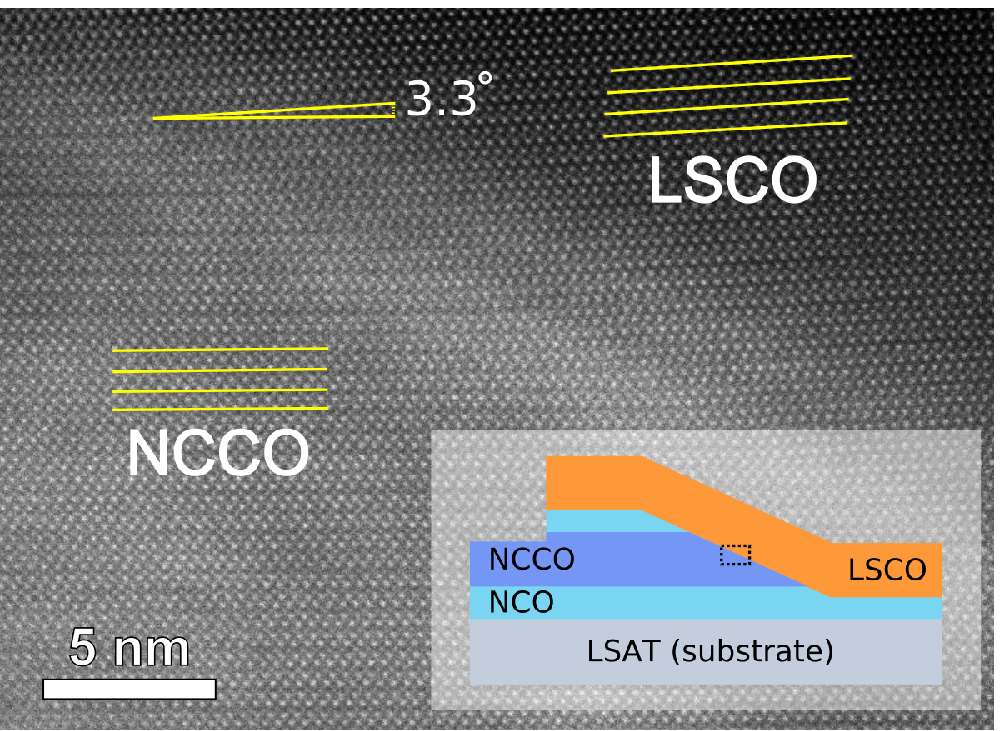}
\caption{HAADF-STEM characterization of the LSCO/NCCO ramp-edge junctions. The inset shows a schematic cross-section of the device, the bottom electrode consists of \SI{70}{\nm} \NCO (NCO), \SI{150}{\nm} \NCCOopt (NCCO) and \SI{20}{\nm} NCO, the top electrode consists of \SI{150}{\nm} \LSCOopt (LSCO). %
The HAADF-STEM image shows a close-up of the ramp area (indicated by the dotted box in the inset), the NCCO and LSCO \textit{ab}-planes are indicated by yellow lines. The LSCO top layer is tilted with respect to the NCCO lattice by \SI{3.3}{\degree}. \label{fig:1}}
\end{figure}

\begin{figure*}
\includegraphics[width = 0.85\textwidth]{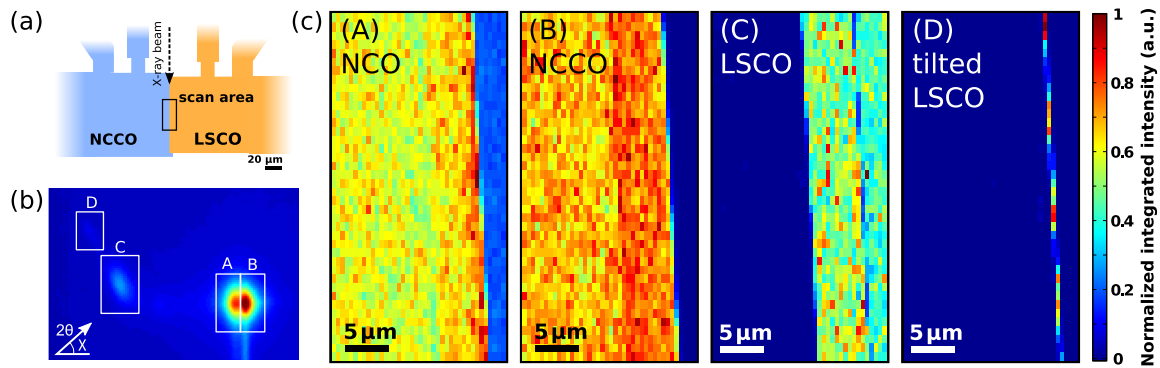}
\caption{Nano-scale X-ray diffraction structural mapping. (a) Schematic top view of the junction with the scan area and the beam direction indicated. (b) Detector image zoomed in around the NCO, NCCO and LSCO \miller{1}{0}{7} peaks. The white boxes are used in the integration for (c). (c) Integrated nXRD intensity maps for NCO \textbf{(A)}, NCCO \textbf{(B)}, LSCO \textbf{(C)} and a tilted phase of LSCO \textbf{(D)}, only visible at the ramp, measured on the scan area indicated in (a). A background is subtracted from all maps and the signal is normalized to the highest intensity pixel. The pixel size is \SI{0.5}{\mum} $\times$ \SI{1}{\mum}, shortest in the horizontal direction.  \label{fig:2}}
\end{figure*}

We have fabricated ramp-edge junctions between superconducting \NCCOopt (NCCO) and \LSCOopt (LSCO) on \LSAT (LSAT) substrates using pulsed laser deposition (PLD), standard photolithography, and ex situ and in situ argon ion milling.  
The samples are grown from commercial and homemade polycrystalline PLD targets. 
The NCCO layer is grown from a target with extra copper added to suppress a parasitic \CNO (CNO) phase \cite{Hoek2014}. 
We use a specifically tailored oxygen annealing and reduction procedure to ensure both layers are superconducting \cite{Hoek2014,Hoek2015a}.
A schematic view of the junction cross-section is shown in the inset of Fig.~\ref{fig:1}. 
The junction consists of a NCCO bottom electrode sandwiched between undoped \NCO (NCO) layers in which a ramp is defined by ex situ argon ion milling under an angle of 45 degrees. 
The NCO layers act as a buffer layer and a capping layer. The LSCO top electrode is deposited after an extra in situ cleaning step of hard and soft argon ion milling. 
The final devices are structured for electronic transport measurements, described elsewhere \cite{Hoek2014a,Hoek2015a}. 
Here, we focus on the structural characterization of the interface between NCCO and LSCO in the ramp area of the junctions. 
We employ two different techniques to probe the interface: scanning nano-focused X-ray diffraction (nXRD) and high-angle annular dark-field scanning transmission electron microscopy (HAADF-STEM). 
The former has become a powerful tool for selective analysis thanks to the development of high brilliance synchrotron sources that employ micro and nano-focused beams \cite{Stangl2009}. 
This allows nXRD to combine sub-micron lateral resolution with high $k$-space resolution \cite{Holt2005,Xiao2005,Hanke2008,Fratini2010,Hrauda2011,Hruszkewycz2011,Klug2011,Poccia2012,Falub2013} for the non destructive study of the crystal structures of buried layers. 
HAADF-STEM complements the nXRD measurements by providing a local cross-section of Z contrast with atomic resolution.
More details on sample fabrication and the HAADF-STEM and nXRD measurement setups can be found in the supplementary information \cite{SupInfo}.

Fig.~\ref{fig:1} shows a HAADF-STEM close-up of the ramp interface.
The NCCO and LSCO layers can be identified, as well as a tilting of the LSCO lattice, indicated with yellow lines. 
The layer composition is verified by electron diffraction and energy dispersive X-ray spectroscopy (EDX), discussed in more detail elsewhere \cite{Hoek2014a,Hoek2015a}.
The exact tilting of the LSCO lattice is determined by looking at the Fourier transforms of the NCCO and the LSCO lattices and is found to be \SI{3.3}{\degree} on the central part of the ramp.

\begin{figure}
\includegraphics[width = 0.9\columnwidth]{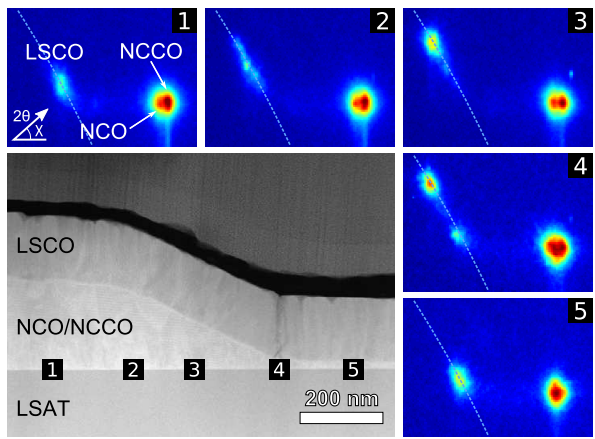}
\caption{LSCO lattice tilting, the bottom left panel shows a low magnification HAADF-STEM cross-section of the sample with five regions indicated. The NCCO and LSCO layers are labeled as well as the LSAT substrate (the NCO layers are not indicated for clarity). The structure is capped with carbon, and electron and ion-deposited Pt layers. The nXRD frames surrounding the HAADF-STEM image show the diffraction pattern for NCO \miller{1}{0}{7}, NCCO \miller{1}{0}{7} and LSCO \miller{1}{0}{7} for the five approximate regions. The diffraction patterns show a gradual change to a tilted phase going from 1 to 3 and an abrupt transition from the tilted phase to the \caxis aligned phase at the grain boundary near region 4. The dashed line follows a constant $2\theta$ angle.  \label{fig:3}}
\end{figure}

Next, we use nXRD to show that the tilting of the LSCO lattice is not a local effect, but occurs along the entire ramp.
Fig.~\ref{fig:2} summarizes the nXRD results. 
A CCD camera collects two dimensional X-ray spectra in a grid scan across the junction interface with the beam aligned to NCCO \miller{0}{0}{8}, which corresponds to a beam angle of around 16 degrees. 
The images in Fig.~\ref{fig:2}(b) and Fig.~\ref{fig:3}(1--5) show the diffraction pattern as it appears on the CCD camera. 
The horizontal direction on the detector corresponds with the $l$ Miller index and the vertical direction with the $h$ Miller index. 
Concentric circles around the direct beam are lines of equal $2\theta$, where the inclination with respect to the horizontal axis corresponds to a tilting or a non-zero $\chi$ angle.
Fig.~\ref{fig:2}(a) shows the measurement geometry, the scan area being \SI{20}{\mum}$\times$\SI{40}{\mum}. 
The focus of the beam allows for the collection a broader range of Bragg reflections, including  \miller{1}{0}{7}, which shows a larger and clearer peak separation. 
Fig.~\ref{fig:2}(b), shows a summation of all the frames collected in one scan zoomed in on the \miller{1}{0}{7} reflections of NCO, NCCO and LSCO and the tilted LSCO phase, labeled A--D. 
Fig.~\ref{fig:2}(c) shows intensity maps of the scan area by integrating over the white boxes labeled A--D in Fig.~\ref{fig:2}(b). 
For the mapping, a constant background is subtracted and the images are normalized to the highest intensity pixel, more details can be found in the supplementary information \cite{SupInfo}. 
The pixel size is \SI{0.5}{\mum} $\times$ \SI{1}{\mum}, with the shortest size in the horizontal direction; each panel is constructed from 1600 diffraction patterns. 
In Fig.~\ref{fig:2}(c) the device architecture of Fig.~\ref{fig:1} can be identified. 
For NCO, panel A in Fig.~\ref{fig:2}(c), going from left to right, first the NCO layer underneath the NCCO electrode is imaged. 
Then the junction overlap area, defined by the complete NCO/NCCO/NCO/LSCO stack, is visible; it has a higher intensity, because here the NCO capping layer has not been etched away (see the inset of Fig.~\ref{fig:1}). 
The ramp is identified as the step-like intensity change from orange to light blue in 1--2 pixels, comparable to the width of the ramp ($\sim$\SI{300}{\nm}). 
Lastly, beyond the ramp on the right, finite intensity remains as the etching process of the ramp is stopped in the NCO layer. 
In the mapping of NCCO, panel B, both the overlap area and the ramp can be identified. 
The overlap area shows a higher intensity because the NCCO outside the overlap area is etched away slightly during the definition of the LSCO contact and the removal of the NCO capping layer. 
Beyond the ramp, only background intensity remains. 
For LSCO, panel C, we get a complementary picture; the overlap area can be identified and no intensity remains to the left of the overlap area, where all the LSCO has been etched away. 
The slightly higher intensity for the overlap area can be explained by a higher crystal quality of LSCO on the overlap as compared to the LSCO grown on the etched surface of NCO on the right side of the ramp. 
At the position where the ramp can be identified in the NCO and NCCO maps, we observe missing intensity in the LSCO map. 
Part of the LSCO \miller{1}{0}{7} intensity shifts on the detector. 
Panel D shows a mapping of the intensity of this shifted phase. 
Here, the intensity along the ramp corresponds to the missing intensity in panel C. 
We observe the same shift for all peaks attributed to LSCO, i.e.  \miller{0}{0}{8}, \miller{1}{0}{7} and \miller{$\bar{\mbox{1}}$}{0}{5}, indicating that the effect is not caused by a rotation of the sample with respect to the beam, see the supplementary information \cite{SupInfo}.
In the shift, the $2\theta$ angle does not change, 
this excludes lattice deformation as the origin of the shift, since that would be accompanied by a change in \caxis length. 
We therefore attribute the shift of the LSCO peaks to a tilting of the LSCO lattice, while the LSCO unit cell remains unchanged. 
The tilt is measured to be around \SI{3}{\degree} from the shift of the \miller{0}{0}{8}, \miller{1}{0}{7} and \miller{$\bar{\mbox{1}}$}{0}{5} peaks of LSCO, see Fig.~\ref{fig:S2} in the supplementary information \cite{SupInfo}.
This is close to the \SI{3.3}{\degree} tilting measured using HAADF-STEM, see Fig.~\ref{fig:1}.

In the grid scan for Fig.~\ref{fig:2}(c), the beam crosses the ramp at a small angle, which allows us to image effects on different parts of the ramp despite the spot size being comparable to the size of the ramp.
Fig.~\ref{fig:3} visualizes the tilting of the LSCO lattice as the X-ray beam is scanned across the junction. 
Panels 1 to 5 in Fig.~\ref{fig:3} show the \miller{1}{0}{7} diffraction peaks for NCO, NCCO and LSCO corresponding to different areas of the ramp-edge structure, their approximate location indicated by the numbers 1--5 the HAADF-STEM image. 
The dashed line in the detector images is a section of a circle of constant $2\theta$ angle, defined by the direct beam center and the main LSCO \miller{1}{0}{7} peak.
The LSCO \miller{1}{0}{7} peak can be seen to shift in different ways at the top and the bottom of the ramp. 
Panel 2 corresponds to the top of the ramp structure where there a gradual change in the ramp angle results in a spread-out peak structure.
On the ramp, only the \SI{3.3}{\degree} tilted phase is observed (panel 3). 
At the bottom of the ramp, the tilted phase meets the \caxis aligned phase in a grain boundary. 
This is reflected in panel 4, where two distinct peaks are observed, with only small streaking between the two. 
Finally, in panel 5, just as in panel 1, the LSCO is fully \caxis aligned again. 

\begin{figure*}
\includegraphics[width = 0.85\textwidth]{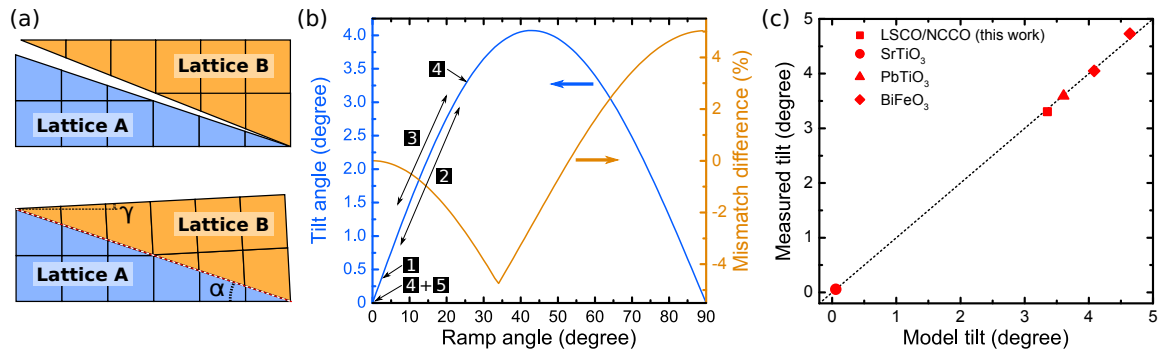}
\caption{Strain accommodation model for lattice tilt on a ramp interface. (a) Schematic representation of the tilting process for two lattices A and B with an in-plane lattice mismatch and different $c/a$ ratios. The ramp angle $\alpha$ and the tilt angle $\gamma$ are indicated. (b) Modelled tilt of the LSCO phase on the NCCO ramp toward the top of the ramp as a function of ramp angle following Eq.~\eqref{eq:tilt} using the measured $a_N/a_L$ and $c_L/c_N$ ratios (blue curve, left axis). The approximate location of the tilt in  regions 1--5 of Fig.~\ref{fig:3} is indicated, where 2 and 3 cover a range of tilt angles.  The other curve shows the absolute lattice mismatch difference between the tilted phase and a \caxis oriented phase, where a negative value indicates a lower mismatch for the tilted phase (orange curve, right axis) (c) Experimental tilt versus model tilt using Eq.~\eqref{eq:tilt} for various oxide systems. The data points for \STO, PbTiO\subs{3} and BiFeO\subs{3} are determined using information from the main text and the TEM measurements in refs\cite{Honig2013,Catalan2011, Zeches2009}. The dotted line indicates the predicted tilt. \label{fig:4}}
\end{figure*}

The tilting observed in our experiment is different from graphoepitaxy that is known to occur in ramp-edge structures of cuprate superconductors\cite{Faley2013a}. 
In graphoepitaxy, \YBCO (YBCO) for example can grow on a ramp etched into a MgO substrate, where the YBCO tilting follows the ramp angle, creating Josephson junctions at the top and bottom of the ramp \cite{Streiffer1990,Faley2013a}. 
For our junctions, the tilt does not follow the ramp angle directly, but is much smaller than the ramp angle. 
We argue that the tilting of the LSCO lattice on the ramp is induced by strain at the interface. 
The tilted phase has a better lattice match along the NCCO facets exposed on the ramp than a \caxis aligned phase. 
Tilting to accommodate lattice mismatch is observed in semiconductor heterostructures for large lattice mismatches in combinations like MnAs/GaAs, GaN/GaAs, GaAs/Si, Cu/GaAs or $\alpha$-Si\subs{3}N\subs{4}/Si \cite{Dodson1988, Riesz1995, Yamada2001, Wan2007}. 
Here, the tilt is determined by the lattice mismatch and the vicinal angle of the substrate. 
In our case, the tilting is determined by the angle of the ramp. 
From HAADF-STEM and atomic force microscopy, we measure a ramp angle of around \SI{26}{\degree}, the facet closest to this angle is NCCO \miller{3}{0}{19}, using literature values for NCCO \cite{Kimura2005}. 
The corresponding LSCO \miller{3}{0}{19} plane has an incline of about \SI{29}{\degree} \cite{Palmer1997}, which gives a \SI{3}{\degree} incline for the \miller{0}{0}{1} planes, when the unit cell is tilted to make the \miller{3}{0}{19} planes of NCCO and LSCO parallel. 
The lattice mismatch between LSCO \miller{3}{0}{19} and NCCO \miller{3}{0}{19} along the ramp is \SI{1.7}{\percent} versus an in-plane mismatch of \SI{4.4}{\percent} for \caxis aligned growth. 
Other planes like \miller{1}{0}{6}, \miller{1}{0}{7}, \miller{2}{0}{13} and \miller{3}{0}{20} also have an incline close to \SI{26}{\degree}. 
For all these planes the corresponding LSCO plane is tilted by about \SI{3}{\degree} and the lattice mismatch is always less than \SI{2}{\percent}. 

In Fig.~\ref{fig:3} we have seen that a gradual change of the ramp angle at the top of the ramp leads to a range of tilt angles for the tilted LSCO phase.
The changing ramp angle exposes different facets of NCCO, which results in LSCO layers with a different tilt. 
Assuming that for each angle the Miller indices of the aligned NCCO and LSCO planes are the same, the tilt angle $\gamma$ is given by
\begin{equation}
\gamma = \arctan\left(\frac{a_Nc_L}{c_Na_L}\tan{\alpha}\right)-\alpha,
\label{eq:tilt}
\end{equation}
where $\alpha$ is the ramp angle in degrees and $a(c)_{N,L}$ is the $a$($c$)-axis length of NCCO (N) and LSCO (L). 
A full derivation can be found in the supplementary information \cite{SupInfo}. 
We can extract the ratios $a_N/a_L$ and $c_L/c_N$ in Eq.~\eqref{eq:tilt} from the HAADF-STEM image of Fig.~\ref{fig:1}. 
By focusing on the ratios we can minimize errors due to image distortion.
We find $a_N/a_L = 1.05$ and $c_L/c_N = 1.098$, to be compared to 1.046 and 1.095, respectively, when using the literature values\cite{Kimura2005,Palmer1997}. 
Fig.~\ref{fig:4}(a) schematically shows the tilting process and in Fig.~\ref{fig:4}(b) the blue curve (left axis) shows the dependence of Eq.~\eqref{eq:tilt} using the measured ratios. 
The orange curve (right axis) shows the difference between the absolute lattice mismatch for the tilted phase (see supplementary information \cite{SupInfo}) and the in-plane absolute lattice mismatch for the \caxis-aligned phase ($|a_N-a_L|/a_N$); a negative value indicates that the tilted phase has a favorable, lower lattice mismatch.

For a ramp angle of \SI{26}{\degree}(\SI{1}{\degree}), we predict a tilt of \SI{3.35}{\degree}(\SI{0.1}{\degree}), where the error bar is mostly generated by inaccuracy in determining the ramp angle. This value falls within the measurement error bar for the measured \SI{3.3}{\degree} tilt.
The tilt ranges associated with panels 1--5 in Fig.~\ref{fig:3} are schematically indicated in Fig.~\ref{fig:4}(b). 
This leads us to conclude that the facet matching model can qualitatively describe the LSCO tilt over the whole ramp structure and quantitatively predict the tilt angle for the dominant ramp angle.

It is interesting to see what Eq.~\eqref{eq:tilt} predicts for other materials commonly used in ramp-edge junctions. 
Most junction designs work with YBCO, which has a $c/a$ ratio very comparable to the 
electron doped cuprates. 
In the configuration of YBCO/NCCO \cite{Takeuchi1995} or for NCCO as interlayer in YBCO--YBCO Josephson junctions \cite{Alff1996}, the predicted tilt is small, <\SI{0.3}{\degree}, and it results in an unfavorable lattice mismatch compared to \caxis aligned growth. 
The same holds for other common interlayer materials like PBCO \cite{Gao1991} and for YBCO grown on ramps etched into substrates like LaAlO\subs{3} \cite{Wen1995} or MgO \cite{Faley2013a}, where in the latter case graphoepitaxy is found due to the large in-plane lattice mismatch. 
A scenario similar to LSCO/NCCO is found in the combination of YBCO and LSCO. Our model predicts a tilt of \SI{3.45}{\degree} for LSCO/YBCO and \SI{-3.16}{\degree} for YBCO/LSCO for a ramp angle of \SI{26}{\degree}. 
These material combinations have been studied by the Maeda group \cite{Fujimaki2006,Gomez2006}, but no reports on the structure are available. 

We note that lattice tilting is also observed across grain boundaries in other oxide systems, for example across a twin grain boundary in tetragonal \STO \cite{Honig2013} or PbTiO\subs{3} \cite{Catalan2011}. 
In both cases the tilting can be described using a simplified version of Eq.~\eqref{eq:tilt}, since across the domain wall $a$ and $c$ are exchanged and the `ramp' angle can be defined by $\arctan\left(a/c\right)$, leading to $\gamma = 2 \arctan(c/a)-90^{\circ}$. 
A tilted lattice also appears across a grain boundary between rhombohedrally distorted ($R$) and tetragonally distorted ($T$) BiFeO\subs{3} for specific substrate strain \cite{Zeches2009, Zhang2011,Infante2011}. 
Here, Eq.~\eqref{eq:tilt} can be used directly by using the pseudocubic $c/a$ ratios associated with the $R$ and the $T$-phase and taking the inclination of the grain boundary as the ramp angle. 
Fig.~\ref{fig:3}(c) shows the measured lattice tilt versus the predicted lattice tilt using Eq.~\eqref{eq:tilt} for the three materials described above and the LSCO/NCCO system described in this Letter. 
The measured tilt and the model tilt are determined using the main text and the TEM figures in references \cite{Honig2013,Catalan2011, Zeches2009}. 
The two data points for BiFeO\subs{3} correspond to the $T/R$ and the $R/T$ configurations. 
The dotted line indicates the predicted tilt; it is clear that the lattice tilt in all four systems can be well described with the facet matching model for strain accommodation. 
Conversely, it also means that the LSCO/NCCO ramp-edge junctions have a crystalline grain boundary contact at the interface. 
Firstly, this is important for the electronic measurements discussed elsewhere \cite{Hoek2014a,Hoek2015a}, but secondly, it also means that a ramp-edge structure can potentially be used to tailor specific, artificial grain boundaries in piezoelectric materials like BiFeO\subs{3} to enhance their piezoelectric properties.
A periodic mesa structure in \STO with ramps on both sides could yield an artificial realization of the mixed phase state as observed by \mbox{Zeches \etal \cite{Zeches2009}}.

In summary, nano-focused XRD has allowed us to identify a tilted phase of LSCO in LSCO/NCCO ramp-edge junctions, also confirmed by HAADF-STEM.
We argue that the origin of the tilting is an interplay between lattice strain and the ramp angle, promoting the LSCO to nucleate in a tilted phase on the exposed NCCO facets at the ramp interface. 
Our facets matching model successfully predicts this behavior for material combinations that have a sufficiently large in-plane lattice mismatch and have a different $c/a$ ratio, with a potential application in realizing artificial domain wall structures in piezoelectric materials.

\begin{acknowledgements}
This research was supported by the Dutch NWO foundation through a VICI grant. 
XRW is supported by an NWO Rubicon grant (2011, 680-50-1114). 
NP is supported by a Marie Curie grant. 
The nXRD experiments were performed on the ID13 beamline at the European Synchrotron Radiation Facility (ESRF), Grenoble, France;  we are grateful to M Burghammer, E Di Cola and G Campi for providing assistance in using beamline ID13. 
Part of the research leading to these results has received funding from the \mbox{European} Union Seventh Framework Programme under Grant Agreement 312483 - \mbox{ESTEEM2} (Integrated Infrastructure Initiative-I3). 
XK and GvT acknowledge funding from the European Research Council under the Seventh Framework Program
(FP7) ERC grant 246791-COUNTATOMS. 
The authors thank S Harkema for valuable discussion.
\end{acknowledgements}


%

\newpage
\section{Supplementary Materials}

\setcounter{figure}{0}
\renewcommand{\thefigure}{S\arabic{figure}}

\begin{figure*}[t]
	\includegraphics[width=\textwidth]{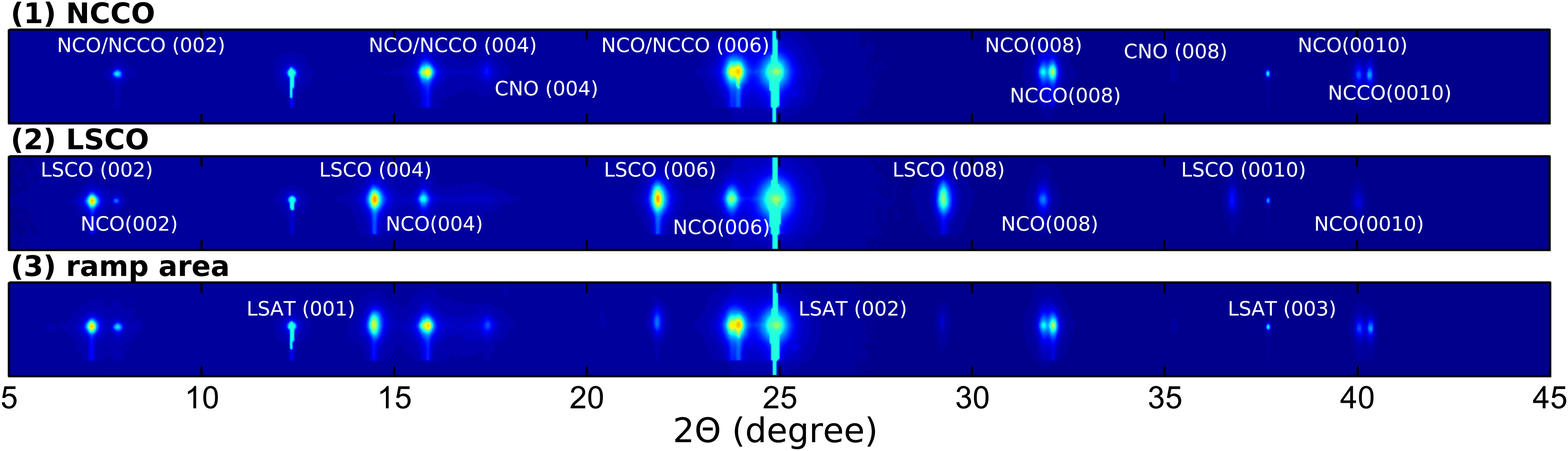}
		\caption{$\theta-2\theta$ nXRD scans on different parts of the device, namely the NCCO layer (1), the LSCO layer (2) and the ramp area (3). Peaks for LSCO, NCCO, NCO, \CNO and LSAT are indicated. The color scale is logarithmic.}
		\label{fig:S1}
\end{figure*}

\subsection{Sample fabrication}
For HAADF-STEM, the bottom layer is structured into \SI{100}{\micro\meter} wide lines and the top layer is not structured to allow for easy sample preparation using focused ion beam (FIB). 
The sample used for nXRD is structured into multiple devices by photolithography and ion milling. 
Sputter deposited Au/Ti contacts are added as the electrodes. 
The outline of a typical device is depicted in Fig.~\ref{fig:2}(a) in the main text. 
These devices, and other similar devices have been characterized electronically and are described elsewhere \cite{Hoek2014a,Hoek2015a}.

\subsection{HAADF-STEM}
Before FIB sample preparation for HAADF-STEM, the sample is covered with a thin amorphous carbon layer, an electron-beam deposited Pt layer and an ion-beam deposited Pt layer. 
A 60--\SI{100}{\nm} thin lamella is defined in the ramp structure using FIB. 
The FIB sample preparation was finished with a final milling and cleaning of the surface from 5\,kV down to 2\,kV, and finally 1\,kV, to minimize damage and unwanted redeposition. 
The HAADF-STEM measurements were performed using a FEI Titan 80--300 ``cubed'' microscope fitted with an aberration-corrector for the probe forming lens and SuperX EDX detectors, operated at \SI{200}{\kilo\volt}. 
High resolution HAADF-STEM images were taken from the ramp interface of the ramp-edge junctions with the substrate tilted to the \millerx{1}{0}{0} zone-axis. 

\subsection{Scanning nano-focused X-ray diffraction}
The nXRD experiments were performed at the ID13 beamline of the European Synchrotron Radiation Facility (ESRF), Grenoble, France. 
The ID13 nanobranch is specialized in the delivery of nano-focused X-ray beams for diffraction experiments. 
The photon source, an \SI{18}{\mm} period in-vacuum undulator, works in the range 5--\SI{17}{\kilo\eV} with the storage ring operating at \SI{6.03}{\giga\eV} in the uniform mode with a current of \SI{200}{\milli\ampere}. 
The beamline uses a Si-111 channel cut crystal monochromator cooled with liquid nitrogen. 
A monochromatic X-ray beam of photon energy \SI{14.9}{\kilo\eV} ($\Delta E/E = 10^{-4}$) was used, which was focused by Kirkpatrick-Baez (KB) mirrors to a \SI{300}{\nm} spot size on the sample (full width at half maximum).
A 16-bit 2D Fast Readout Low Noise charged coupled device (FReLoN CCD) detector with $2048 \times 2048$ pixels of \SI{51}{\micro\meter} $\times$ \SI{51}{\micro\meter} was used. 
The detector was placed \SI{50}{\mm} behind the sample and offset. 
Diffraction images were obtained after correcting the 2D images for dark noise, flat field and distortion. 
The FreLoN CCD camera records the intensity of the peak that correspond to the square root volume of the domains in the selected sample surface spot. 
The intensity is integrated over square subareas of the images recorded by the FreLoN CCD detector in reciprocal-lattice units of the main crystalline reflections at each point ({\em x,y}) of the sample reached by the translator. 
For most of the experiments discussed here, the angle of incidence of the X-ray beam with the sample was around \SI{16}{\degree}. 
This gives an effective interaction length of \SI{300}{\nano\meter} perpendicular to the beam and 0.8--\SI{1.3}{\mum} in the beam direction due to the finite thickness of the sample, as the thickness of the layers of interest is 150--\SI{300}{\nano\meter}. 

In the experiment, the device is first aligned by microscope and then $\theta-2\theta$ scans are performed on selected areas by rotating the sample in the beam with the CCD detector stationary; for each angle step a CCD frame is recorded.
The \thth map is constructed by summing all the frames. 
Antisymmetric reflections are removed in the image processing by only collecting in a moving box that follows $2\theta$.
Fig. \ref{fig:S1}(a) shows \thth scans (color scale is logarithmic) for the NCCO electrode (\textit{1}), the LSCO electrode (\textit{2}) and the ramp area (\textit{3}). 
The \miller{0}{0}{$l$} reflections, with $l$ even, for LSAT, NCCO, NCO and LSCO are indicated. 
The parasitic \CNO phase is also present in all three figures, since there is always a layer of NCO remaining. 
The \CNO \miller{0}{0}{4} and \miller{0}{0}{8} reflections can be identified. 
Strong film peaks and especially the LSAT \miller{0}{0}{1} and \miller{0}{0}{2} substrate peaks show streaks due to saturation of the CCD detector.

Next, the sample is aligned to NCCO \miller{0}{0}{8} for the diffraction mapping. 
NCCO \miller{0}{0}{8} is selected over the higher intensity peak of \miller{0}{0}{6}, because at this angle NCO \miller{0}{0}{8} and NCCO \miller{0}{0}{8} can be separately resolved, see Fig.~\ref{fig:S1}. 
Furthermore, no nearby substrate peaks give a changing background due to e.g. mosaic spread in the substrate and saturation of the CCD detector. 
As shown in Fig.~\ref{fig:2}(c), the sample is scanned in a grid across the junction area with a step size of \SI{1}{\mum} along the junction and \SI{0.5}{\mum} perpendicular to the ramp with the junction interface parallel to the beam to get the highest resolution across the junction. 
At each pixel, a CCD image is collected. 
A beam convergence, due to the focusing of the beam, enables us to measure the \miller{0}{0}{8}, \miller{1}{0}{7} and \miller{$\bar{\mbox{1}}$}{0}{5} reflections of NCCO, NCO and LSCO at the same time, allowing for a mapping of all the materials of the device. 

For the mappings of Fig.~\ref{fig:2}(c), the intensity is integrated over the boxes indicated in Fig.~\ref{fig:2}(b) for each pixel.
A background is subtracted by using the average intensity level when no peak is present, i.e. the dark blue areas in panels B--D in Fig.~\ref{fig:2}(c). 
For panel A in Fig.~\ref{fig:2}(c), the background of panel B is used, scaled to the size of the integration box. 
Each image is then normalized to the highest intensity pixel.

\begin{figure}
\includegraphics[width=\columnwidth]{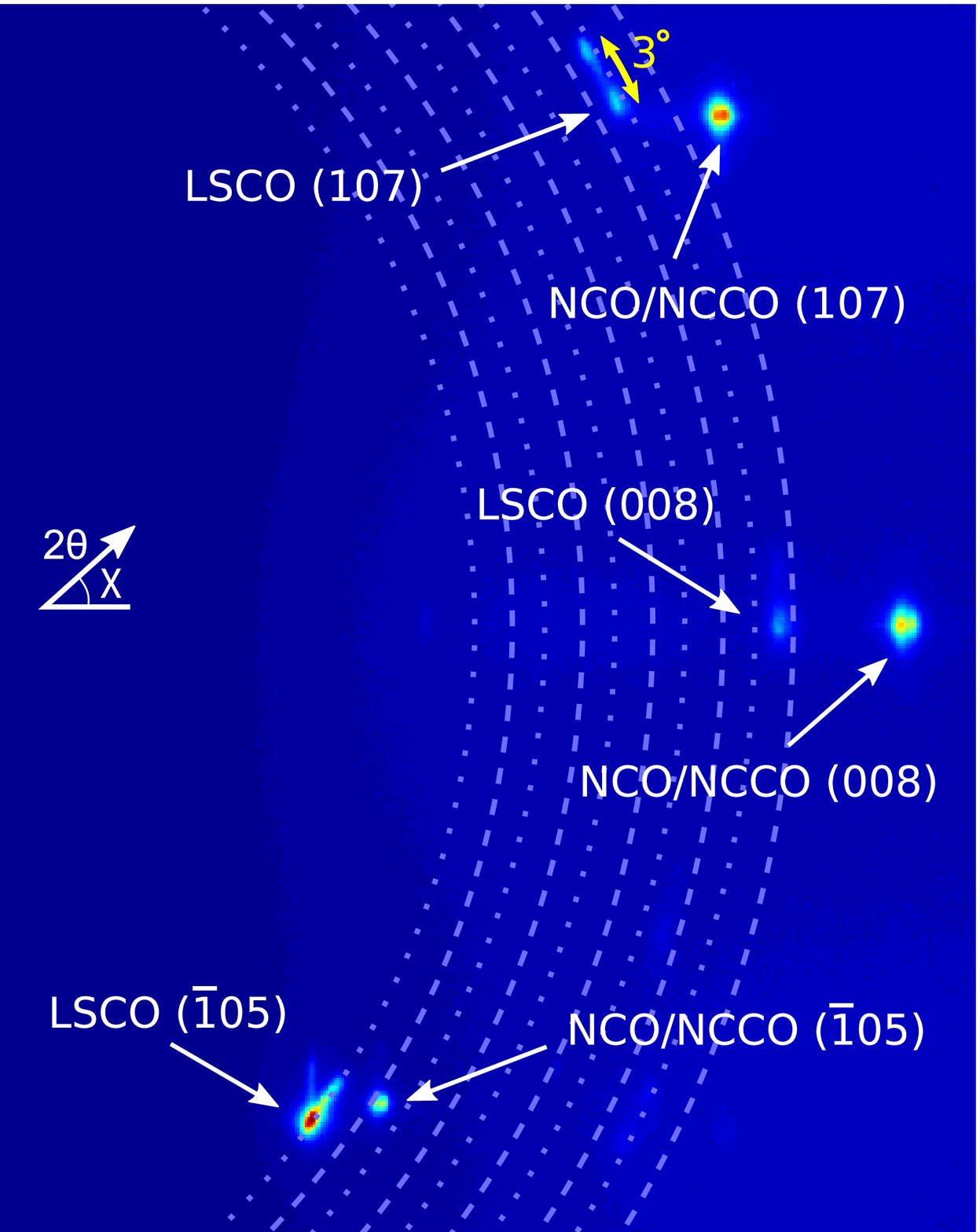}
\caption{Summation of all CCD frames collected in the mapping of Fig.~\ref{fig:2} (logarithmic color scale), showing that the shifted LSCO phase tilts without changing $2\theta$. NCO/NCCO and LSCO \miller{0}{0}{8}, \miller{1}{0}{7} and \miller{$\bar{\mbox{1}}$}{0}{5} are labeled. The dotted lines indicate equal $2\theta$ angles.}
\label{fig:S2}
\end{figure}

In the main text, we show mappings of the \miller{1}{0}{7}, where the tilting of the LSCO lattice is most visible. 
Mapping of the other available peaks shows the same result, but with cross-talk between the different maps due to overlapping integration areas.
Fig.~\ref{fig:S2} shows a summation of all the frames collected during the mapping on a logarithmic scale.
Here, it can be seen that the tilting of the LSCO lattice is observed for the \miller{0}{0}{8}, \miller{1}{0}{7} and \miller{$\bar{\mbox{1}}$}{0}{5}.
The difference in visibility between e.g. LSCO \miller{1}{0}{7} and \miller{$\bar{\mbox{1}}$}{0}{7} is caused by a small, in-plane rotation of the sample. 

\subsection{Lattice tilt}
\label{app:tilt}
For the derivation of Eq.~\eqref{eq:tilt} in the main text, we assume that the tilting originates from a matching of the same \miller{$h$}{0}{$l$} planes of the two materials. We will use the schematic diagram in Fig.~\ref{fig:4}(a) for reference. The \miller{$h$}{0}{$l$} facet of `lattice A' will make an angle $\alpha$ with the ($a,b$)-plane, which is also the ramp angle. The same facet in `lattice B' will make a different angle $\alpha^{\prime}$, in this case larger than $\alpha$. Both angles can be expressed using the lattice parameters of the two lattices,
\begin{equation}
\tan\alpha(\alpha^{\prime}) = \frac{n c_{A(B)}}{m a_{A(B)}},
\label{eq:tanratio}
\end{equation}
where $n$ and $m$ are the number of unit cells that define the angle associated with the \miller{$h$}{0}{$l$} plane and $a$ and $c$ are the in and out-of-plane lattice parameters. The tilt angle can now be defined as $\gamma = \alpha^{\prime}-\alpha$, where $\alpha^{\prime}$ can be eliminated using
\begin{equation}
\frac{n}{m} = \frac{a_A}{c_A}\tan\alpha,
\end{equation} 
to get
\begin{equation}
\tan\alpha^{\prime} = \frac{a_A c_B}{c_A a_B}\tan\alpha\,;
\label{eq:elim}
\end{equation}
which gives for $\gamma$,
\begin{equation}
\gamma = \arctan\left(\frac{a_Ac_B}{c_Aa_B}\tan{\alpha}\right)-\alpha. 
\end{equation}

These equations can also be used to get a figure for the lattice mismatch between the tilted planes. We associate a length $L$($L^{\prime}$) with the \miller{$h$}{0}{$l$} facet of lattice A(B),
\begin{equation}
L(L^{\prime}) = \frac{m a_{A(B)}}{\cos \alpha (\alpha^{\prime})}.
\end{equation}
This we use to express a lattice mismatch
\begin{equation}
\frac{L-L^{\prime}}{L} = 1 - \frac{a_B \cos \alpha}{a_A \cos \alpha^{\prime}},
\end{equation}
where Eq.~\eqref{eq:elim} can be used again to express $\cos \alpha^{\prime}$ in terms of $\alpha$ and the lattice parameters of the two materials.

\end{document}